\documentclass[prd,twocolumn,eqsecnum,longbibliography,nofootinbib]{revtex4-1}

\usepackage{bm}
\usepackage[colorlinks=true,urlcolor=blue,anchorcolor=blue,citecolor=blue,linkcolor=blue,pagecolor=blue,linktocpage=true,pdfa=true]{hyperref}
\usepackage{amsthm,amsmath,latexsym,amssymb,mathrsfs,bm}
\usepackage{graphicx}
\newtheoremstyle{dotless}{}{}{\itshape}{}{\bfseries}{}{ }{}
\theoremstyle{dotless}
\newtheorem*{thm}{}

\begin{document}
\title{Maximal extensions and singularities in inflationary spacetimes}
\author{Daisuke Yoshida}
\email{d.yoshida@physics.mcgill.ca}
\affiliation{Department of Physics, McGill University, Montr\'eal, QC, H3A 2T8, Canada}
\author{Jerome Quintin}
\email{jquintin@physics.mcgill.ca}
\thanks{Vanier Canada Graduate Scholar}
\affiliation{Department of Physics, McGill University, Montr\'eal, QC, H3A 2T8, Canada}

\begin{abstract}
Extendibility of inflationary spacetimes with flat spatial geometry is investigated.
We find that the past boundary of an inflationary spacetime becomes a so-called parallely propagated curvature singularity
if the ratio $\dot{H}/a^2$ diverges at the boundary, where $\dot{H}$ and $a$ represent the time derivative of
the Hubble parameter and the scale factor, respectively.
On the other hand, if the ratio $\dot{H}/a^2$ converges, then the past boundary is regular and continuously extendible.
We also develop a method to judge the continuous ($C^0$)
extendibility of spacetime in the case of slow-roll inflation driven by a canonical scalar field.
As applications of this method, we find that Starobinsky inflation has a $C^0$ parallely propagated curvature singularity,
but a small field inflation model with a Higgs-like potential does not.
We also find that an inflationary solution in a modified gravity theory with limited curvature invariants is free of such a singularity
and is smoothly extendible.
\end{abstract}

\maketitle

\section{Introduction}
Spacetime singularities remain deep mysteries in gravitational theories.
According to the singularity theorems of Penrose and Hawking \cite{Penrose:1964wq,Hawking:1967ju,Hawking:1973uf},
the occurrence of singularities is inevitable in General Relativity when the stress-energy tensor satisfies some energy conditions.
For example, the singularity theorems ensure the presence of the initial Big Bang singularity in cosmology,
also known as the initial singularity problem, which is characterized by incomplete timelike geodesics,
provided the strong energy condition is satisfied. 

Inflation \cite{Guth:1980zm} is a well-studied structure formation scenario for the very early universe, and it is supported by recent observations of
the cosmic microwave background anisotropies \cite{Ade:2015xua,Ade:2015lrj}. Since inflationary cosmology violates the strong energy condition,
it was expected to solve the initial singularity problem. For example, Starobinsky's inflationary model \cite{Starobinsky:1980te}
was originally proposed as a possible resolution to the initial singularity problem.
However, even though the strong energy condition is violated,
it does not guaranty the absence of a singularity.
In fact, Borde and Vilenkin \cite{Borde:1993xh,Borde:1996pt} showed that eternal inflation\footnote{Strictly speaking,
the theorem was shown under the assumption that the volume of the past of an inflationary region is finite (assumption D in Ref.~\cite{Borde:1996pt})
rather than the assumption of eternal inflation itself. However, this assumption is naturally satisfied in
eternal inflation models based on the old inflation scenario as explained in Ref.~\cite{Borde:1996pt}.}
models are null-geodesically incomplete to the past provided the null energy condition is satisfied.
After that, another interesting theorem was shown by Borde, Guth and Vilenkin \cite{Borde:2001nh}.
They generalized the concept of Hubble parameter to a general (inhomogeneous and anisotropic)
spacetime where comoving geodesic congruence is defined. Then, they showed that a geodesic is
incomplete if the averaged Hubble parameter along this geodesic is positive.
Interestingly, neither the null energy condition nor eternal inflation is assumed, and the theorem can be applied to
a very wide class of inflationary models. This result motivates the exploration of alternative very early universe scenarios
that could evade the assumptions of the theorem to be past complete
(see, e.g., Refs.~\cite{Mukhanov:1991zn,Brandenberger:1993ef,Yoshida:2017swb,Creminelli:2016zwa} and references therein).

Let us quickly review the analysis of Ref.~\cite{Borde:2001nh} with an emphasis on the case of a flat Friedmann-Lema\^{i}tre-Robertson-Walker (FLRW)
spacetime. In this case, the averaged Hubble parameter along a null geodesic is defined by
\begin{equation}
 H_{\mathrm{av}}[t_f, t_i] \equiv \frac{1}{\lambda(t_f) - \lambda(t_{i})} \int^{\lambda(t_f)}_{\lambda(t_i)} \mathrm{d} \lambda \, H(\lambda) \, ,
\end{equation}
where $\lambda(t)$ is an affine parameter of a null geodesic at time $t$, and $H$ is the Hubble parameter.
One can show that the integral on the right-hand side is smaller or equal
than unity by suitably choosing the affine parameter (see Ref.~\cite{Borde:2001nh}).
Thus, if $H_{\mathrm{av}}[t_f,-\infty]>0$, one obtains
\begin{align}
 0 < &~\lim_{t_i\rightarrow -\infty} \frac{1}{\lambda(t_f) - \lambda(t_{i})} \int^{\lambda(t_f)}_{\lambda(t_i)}
  \mathrm{d} \lambda \, H(\lambda) \nonumber \\
 \leq &~\lim_{t_i\rightarrow -\infty} \frac{1}{\lambda(t_f) - \lambda(t_i)}\,.
\end{align} 
This means that the affine parameter $\lambda(t)$ has to be finite in the limit where $t \rightarrow - \infty$, and consequently,
the corresponding flat FLRW spacetime is past incomplete. If the initial stage of the Universe is described by inflation,
then the averaged Hubble parameter must be positive. Therefore, there appears to be no hope of avoiding
the past incompleteness of any inflationary flat FLRW spacetime.

\begin{figure}[pt]
\begin{minipage}{0.9\hsize}
\begin{center}
 \includegraphics[width=0.9\hsize]{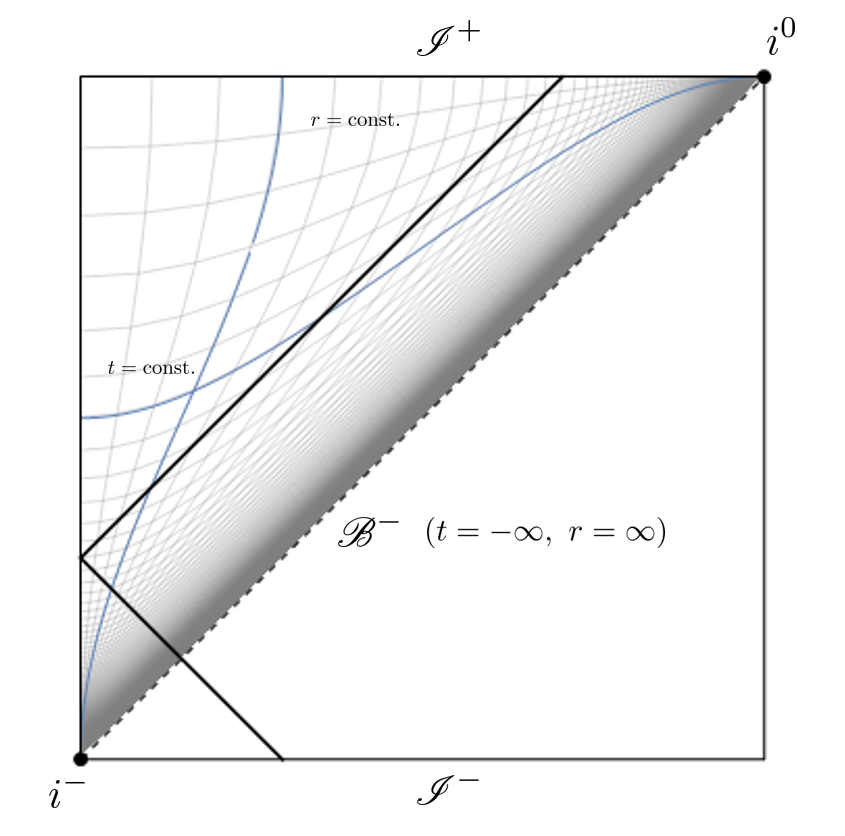}
 \caption{Penrose diagram of de Sitter space:
 The upper half triangle region corresponds to the spacetime region which is covered by the flat FLRW coordinate patch.
 The bold line represents a null geodesic. The null geodesic is incomplete in the flat FLRW patch and reaches the past null
 boundary $\mathscr{B}^{-}$ (represented by the dashed line) at finite affine length.
 However, this geodesic is complete in the entire de Sitter spacetime.
 In the flat FLRW patch, the vertical gray lines represent surfaces of constant radius $r$,
 and the horizontal lines represent surfaces of constant time $t$. Examples are highlighted in blue.
 Also, $i^0$ and $i^-$ denote spatial infinity and past timelike infinity, respectively, while
 $\mathscr{I}^+$ and $\mathscr{I}^-$ denote future and past null infinity, respectively.}
\label{fdS}
\end{center}
\end{minipage}
\end{figure}

Nonetheless, it might be possible to \emph{extend} the flat FLRW spacetime beyond the end points of the incomplete geodesics,
which we call the past boundary $\mathscr{B}^{-}$. The important observation here is that the above discussion can be applied even
when the spacetime is exactly flat de Sitter space. This implies that flat de Sitter space must be past incomplete.
This does \emph{not} contradict the fact that de Sitter space is free of singularities,
because the flat patch of de Sitter space covers only half of the entire de Sitter space (see Fig.~\ref{fdS}).
Thus, even though flat de Sitter space is actually past incomplete, $\mathscr{B}^{-}$ is not singular,
and the spacetime can be extended to the entire nonsingular de Sitter space beyond $\mathscr{B}^{-}$.
In the context of the general setting of Ref.~\cite{Borde:2001nh}, this would happen when the comoving geodesic congruence,
which defines the Hubble parameter, does not cover the entire spacetime.

Since inflationary spacetimes are effectively described by flat de Sitter space,
it is natural to expect that the past incompleteness of inflation in a flat FLRW coordinate patch is just an apparent one
and that there might exist a nonsingular maximal extension of it.
The purpose of this paper is to explore such extendibility of inflationary spacetimes with flat spatial geometry.

The question that must be answered is how one can judge the extendibility of the past boundary $\mathscr{B}^{-}$.
The classification of a boundary of spacetime was studied in \cite{Ellis:1977pj} (see also \cite{Hawking:1973uf}).
In Ref.~\cite{Clarke1973,Clarke1982}, it was shown that a spacetime is locally inextendible (or inextensible) if and only if any component of
the Riemann tensor and its covariant derivatives measured in a parallely propagated (p.\,p.) tetrad basis diverges in the limit to the boundary.
This kind of singularity is called a \emph{p.\,p.\ curvature singularity} \cite{Hawking:1973uf} or simply
a \emph{curvature singularity} \cite{Ellis:1977pj}.
We note that the more common \emph{scalar curvature singularity}, where a scalar curvature invariant blows up,
is necessarily also a p.\,p.\ curvature singularity.
However, the converse is not always true, i.e., it is possible that a p.\,p.\ curvature singularity
may not be a scalar curvature singularity.
This kind of singularity is called an \emph{intermediate singularity} \cite{Ellis:1974ug} or a \emph{nonscalar singularity} \cite{Ellis:1977pj}.
One notes that a locally extendible spacetime also includes a globally inextendible one.
A boundary that can be extended locally but not globally is called a \emph{locally extendible singularity} \cite{Ellis:1974ug} or a
\emph{quasi-regular singularity} \cite{Ellis:1977pj}. A conical singularity and the singularity in Taub-NUT spacetime correspond
to locally extendible singularities (see, e.g., Ref.~\cite{Ellis:1977pj}). A globally extendible boundary is called a \emph{regular boundary},
and there is no obstacle to extend spacetime beyond this boundary. An example of this is the boundary of flat de Sitter space discussed above.
To summarize, the finiteness of the components of the Riemann tensor with respect to a p.\,p.\ tetrad basis
is crucial to discuss the extendibility of spacetimes, at least locally.

Our paper is organized as follows.
In the next section, we construct an affine parametrization of a null geodesic and explicitly demonstrate its past incompleteness
as $t \rightarrow - \infty$. Then, in Sec.~\ref{sec:pptetrad}, we explicitly construct a p.\,p.\ tetrad for flat FLRW spacetimes
and discuss the extendibility of inflationary cosmologies.
We find that a flat FLRW spacetime is continuously inextendible if
the quantity $\dot{H}/a^2$ diverges in the limit $t \rightarrow - \infty$. On the other hand, it is continuously extendible if $\dot{H}/a^2$
is finite in the limit $t \rightarrow - \infty$. In Sec.~\ref{sec:coordinate}, we give a concrete method to extend the spacetime
(when it is shown to be extendible). After that, we give examples of analytic spacetimes in Sec.~\ref{sec:examples}.
There, we demonstrate the maximal extension of each inflationary spacetime or show the presence of the p.\,p.\ singularity explicitly.
In Sec.~\ref{inflationarymodel}, we discuss the implications for inflationary models.
Specifically, we derive the condition for single field slow-roll models to be continuously extendible, and we find that the Starobinsky
model has a continuous p.\,p.\ singularity, but a small field inflation model does not. Moreover, we investigate the absence
of p.\,p.\ curvature singularities in a model of modified gravity studied in Refs.~\cite{Mukhanov:1991zn,Brandenberger:1993ef,Yoshida:2017swb}.
The final section is devoted to the summary and discussion.

\section{Setup and past incompleteness of inflationary spacetimes}

Throughout this paper, we focus on flat FLRW spacetime,
\begin{equation}
 g_{\mu\nu}\mathrm{d}x^{\mu}\mathrm{d}x^{\nu} = - \mathrm{d} t^2 + a(t)^2 \left(\mathrm{d}r^2
  + r^2 \, \mathrm{d} \Omega_{(2)}^2\right)\,, \label{FLRWcoord}
\end{equation}
where $\mathrm{d}\Omega_{(2)}^2=\mathrm{d}\theta^2+\sin^2\theta\,\mathrm{d}\phi^2$ is the metric of the unit 2-sphere.
We assume that the early stage of cosmological evolution is described by inflationary
exponential expansion, which leads to the past boundary $\mathscr{B}^{-}$ as we will see below.
Precisely, we assume that the comoving geodesics (for which the spatial coordinates $r$, $\theta$, and $\phi$ are constants) are past
complete\footnote{We note that extensions of spacetimes where the comoving geodesics are past incomplete,
e.g.~cosmologies with an initial Big Bang singularity, was discussed in Ref.~\cite{Klein:2016tzt}.},
i.e., the comoving time $t$ is defined all the way to $t \rightarrow -\infty$,
and the scale factor $a(t)$ approaches that of de Sitter space in the limit where $t \rightarrow - \infty$. Specifically,
the assumption is that, asymptotically,
\begin{equation}
 a(t) \simeq \bar{a} \, e^{H_{\Lambda} t } \qquad \mathrm{as}\ t \rightarrow -\infty \, , \label{ainf}
\end{equation}
where $\bar{a}$ and $H_{\Lambda}$ are positive constants.
In that limit, $H_\Lambda$ represents the Hubble parameter since $H\equiv\dot a/a\simeq H_\Lambda$.
Throughout this paper, a dot denotes a derivative with respect to physical time $t$.

Under the above assumption, we can directly confirm the incompleteness of a null geodesic.
In order to construct an affine parametrization of a null geodesic, it is useful to introduce the conformal time $\eta$ defined by
\begin{equation}
 \eta(t) \equiv \int^{t} \frac{\mathrm{d}t'}{a(t')}\,.
\end{equation}
Then, it is straightforward to see that a curve parametrized by $\tilde{\lambda}$ as follows,
\begin{equation}
 x^{\mu}(\tilde{\lambda}) = \left(\eta(\tilde{\lambda}), r(\tilde{\lambda}), \theta(\tilde{\lambda}), \phi(\tilde{\lambda}) \right)
 =\left(\tilde{\lambda},-\tilde{\lambda},0,0\right)\,,\label{nullgeotilde}
\end{equation}
is a null geodesic. However, the parameter $\tilde{\lambda}$ is not an affine parameter because the right-hand side of the geodesic equation,
\begin{equation}
 \tilde{k}^{\nu}\nabla_{\nu}\tilde{k}^{\mu} = 2\frac{\partial_{\eta} a}{a}\tilde{k}^{\mu}\,,
\end{equation}
does not vanish. Here, $\tilde{\bm{k}}$ is the tangent vector of the null geodesic \eqref{nullgeotilde},
and it is given by
\begin{equation}
 \tilde{\bm{k}} = \tilde{k}^{\mu}\bm{\partial}_{\mu}
 = \frac{\mathrm{d}x^{\mu}(\tilde{\lambda})}{\mathrm{d}\tilde{\lambda}} \bm{\partial}_{\mu}
 = \bm{\partial}_{\eta} - \bm{\partial}_{r}\,.
\end{equation}
An affine parameter of the null geodesic \eqref{nullgeotilde} can be derived by re-parameterizing $\tilde{\lambda}$.
Let us consider a new parameter $\lambda = \lambda(\tilde{\lambda})$ and its corresponding tangent vector
$k^{\mu} = \mathrm{d} x^{\mu} / \mathrm{d} \lambda$. 
Then, the tangent vector with respect to $\tilde{\lambda}$ can be expressed in terms of $\lambda$ as
\begin{equation}
 \tilde{k}^{\mu} = \frac{\mathrm{d}x^{\mu}}{\mathrm{d}\tilde{\lambda}}
 = \frac{\mathrm{d}\lambda}{\mathrm{d}\tilde{\lambda}} \frac{\mathrm{d}x^{\mu}}{\mathrm{d}\lambda}
 = \frac{\mathrm{d}\lambda}{\mathrm{d}\tilde{\lambda}} k^{\mu}\,,
\end{equation}
and one can derive the geodesic equation for $k^{\mu}$:
\begin{equation}
 k^{\nu} \nabla_{\nu} k^{\mu} = - \frac{1}{(\partial_{\eta}\lambda)^2}\left(\partial_{\eta}^2 \lambda
  - 2 \frac{\partial_{\eta}a}{a} \partial_{\eta}\lambda \right) k^{\mu}\,.\label{geoeq} 
\end{equation}
In order for $\lambda$ to be an affine parameter, the right-hand side of Eq.~\eqref{geoeq} has to vanish. This is only the case if
\begin{equation}
 \mathrm{d} \lambda  \propto a^2 \, \mathrm{d} \eta = a \, \mathrm{d}t\,.
\end{equation} 
Thus, the null geodesic \eqref{nullgeotilde} is past complete if and only if
the affine parameter diverges in the limit where $t \rightarrow -\infty$, i.e., if the integral
\begin{equation}
 \lambda[t_f, -\infty] = \int_{-\infty}^{t_f} \mathrm{d}t~a(t) \label{deflambda}
\end{equation}
is infinite for arbitrary $t_f$.
In the case of inflation, with the assumption \eqref{ainf}, the integral \eqref{deflambda} converges:
\begin{equation}
 \lambda[t_f, -\infty] \simeq \bar{a} \int_{-\infty}^{t_f} \mathrm{d}t~e^{H_{\Lambda} t} = \frac{\bar{a}}{H_{\Lambda}}e^{H_{\Lambda} t_f}\,.
\end{equation}  
Therefore, flat FLRW spacetime with the assumption \eqref{ainf} has incomplete null geodesics,
and there is a past boundary $\mathscr{B}^{-}$.

In the next section, we construct a p.\,p.\ tetrad basis along these incomplete null geodesics and discuss the extendibility.

\section{The parallely propagated curvature singularity in inflationary spacetimes}
\label{sec:pptetrad}

To discuss the local extendibility of an inflationary spacetime, one needs to construct a p.\,p.\ tetrad basis
along the null geodesic from the previous section since that is the basis which is well defined on the boundary $\mathscr{B}^-$.
First, one can construct a simple tetrad $\hat{\bm{e}}^M$, which we call the FLRW tetrad, as follows:
\begin{subequations} 
 \label{FLRWtetrad}
\begin{align}
 &\hat{\bm{e}}^{0} = \mathbf{d}t = a \, \mathbf{d}\eta\,; \\
 &\hat{\bm{e}}^{1} = a \, \mathbf{d}r\,; \\
 &\hat{\bm{e}}^{2} = a \, r \, \mathbf{d}\theta\,; \\
 &\hat{\bm{e}}^{3} = a \, r \sin \theta \, \mathbf{d}\phi\,.
\end{align}
\end{subequations}
Though the FLRW tetrad components are p.\,p.\ along comoving timelike geodesics, they are not parallel
along the null geodesic of the previous section. This can be confirmed by calculating $\bm{\nabla}_{\bm{k}}\hat{\bm{e}}^M$.
Indeed, the covariant derivatives of $\hat{\bm{e}}^{0}$ and $\hat{\bm{e}}^{1}$ along $\bm{k}$ are nonvanishing:
\begin{subequations}
\begin{align}
 k^\mu \nabla_{\mu} \hat{\bm{e}}^{0} &= H \, \mathbf{d} r \, ; \\
 k^\mu \nabla_{\mu} \hat{\bm{e}}^{1} &= H \, \mathbf{d}\eta\,.
\end{align}
\end{subequations}
We note that only $\hat{\bm{e}}^{2}$ and $\hat{\bm{e}}^{3} $ are p.\,p.\ along $\bm{k}$.

Let us now construct a p.\,p.\ tetrad $\bm{e}^M$ simply by parallel transport of the FLRW tetrad at a point
$(\eta, r) = (\eta_0, r_0)$ along the null geodesic.
Since any two tetrad bases are related through a Lorentz transformation, we can write a p.\,p.\ tetrad basis
$\bm{e}^M$ as a Lorentz transformation of the FLRW tetrad,
\begin{subequations}
\begin{align}
 &\bm{e}^{0} = \cosh \zeta(\eta,r) \, \hat{\bm{e}}^{0} + \sinh \zeta(\eta,r) \, \hat{\bm{e}}^{1},\\
 &\bm{e}^{1} = \sinh \zeta(\eta,r) \, \hat{\bm{e}}^{0} + \cosh \zeta(\eta,r) \, \hat{\bm{e}}^{1},
\end{align}
\end{subequations}
with $\bm{e}^{2} = \hat{\bm{e}}^{2}$ and $\bm{e}^{3} = \hat{\bm{e}}^{3}$.
The Lorentz factor (or rapidity) $\zeta$ satisfies $\zeta(\eta_0, r_0) = 0$
so that $\bm{e}^M$ coincides with $\hat{\bm{e}}^M$ at the given point $(\eta_0, r_0)$.
Then, the covariant derivative of $\bm{e}^M$ along $\bm{k}$ can be expressed in terms of the Lorentz factor $\zeta$ as
\begin{subequations}
\label{kde}
\begin{align}
 &k^{\mu}\nabla_{\mu} \bm{e}^{0} = \frac{\partial_{\eta}a + a \left( \partial_{\eta} \zeta - \partial_{r} \zeta \right)}{a^3} \, \bm{e}^{1}\,, \\
 &k^{\mu}\nabla_{\mu} \bm{e}^{1} = \frac{\partial_{\eta}a + a \left( \partial_{\eta} \zeta - \partial_{r} \zeta \right)}{a^3} \, \bm{e}^{0}\,.
\end{align}
\end{subequations}
For the $\bm{e}^M$'s to be parallely propagated, the right-hand sides in Eq.~\eqref{kde} have to vanish.
Thus, one solution for the Lorentz factor $\zeta(\eta, r)$ is
\begin{align}
 \zeta = - \ln\left(\frac{a}{a_0}\right)\,,\label{A}
\end{align}
where $a_0\equiv a(\eta_0)$ is the value of the scale factor at $\eta = \eta_0$.
In the limit where $a \rightarrow 0$, i.e.~toward the boundary $\mathscr{B}^{-}$, the expression \eqref{A} tells us that the
p.\,p.\ basis is obtained by parallel transport through an infinite boost from the FLRW tetrad basis.
Since the p.\,p.\ tetrad is a well-defined basis to discuss the extendibility of the curve beyond $\mathscr{B}^{-}$,
this fact also tells us that the FLRW tetrad basis is ill defined in the limit to the boundary $\mathscr{B}^{-}$.
By plugging \eqref{A} into the definition of $\bm{e}^M$, we obtain concrete expressions for our p.\,p.\ tetrad, 
\begin{subequations}
\begin{align}
 \bm{e}^{0}=&~\frac{1}{2}\left(1 + \frac{a^2}{a_0^2}\right) \frac{a_0}{a} \, \hat{\bm{e}}^{0} + \frac{1}{2} \left(1 - \frac{a^2}{a_0^2} \right) \frac{a_0}{a} \, \hat{\bm{e}}^{1}\,, \\
 \bm{e}^{1}=&~\frac{1}{2}\left(1 - \frac{a^2}{a_0^2}\right) \frac{a_0}{a} \, \hat{\bm{e}}^{0} + \frac{1}{2} \left(1 + \frac{a^2}{a_0^2} \right) \frac{a_0}{a} \, \hat{\bm{e}}^1\,.
\end{align} 
\end{subequations}
Inversely, the original FLRW tetrad can be expressed in terms of the p.\,p.\ tetrad as
\begin{subequations}
\label{FLRWtoPP}
\begin{align}
 \hat{\bm{e}}^0 =&~\frac{1}{2}\left(1 + \frac{a^2}{a_0^2}\right) \frac{a_0}{a} \, \bm{e}^{0} - \frac{1}{2}\left(1 - \frac{a^2}{a_0^2} \right) \frac{a_0}{a} \, \bm{e}^{1}\,, \\
 \hat{\bm{e}}^1 =&-\frac{1}{2}\left(1 - \frac{a^2}{a_0^2}\right) \frac{a_0}{a} \, \bm{e}^{0} + \frac{1}{2}\left(1 + \frac{a^2}{a_0^2} \right) \frac{a_0}{a} \, \bm{e}^{1}\,.
\end{align}
\end{subequations}

As studied in Refs.~\cite{Ellis:1977pj,Clarke1973,Clarke1982,Hawking:1973uf}, components of the Riemann tensor (and derivatives thereof)
are crucial to discuss the local extendibility of a boundary. Since flat FLRW spacetime is conformally flat, any independent components
of the Riemann tensor is described by the Ricci tensor. In the FLRW tetrad basis $\hat{\bm{e}}^M$, the Ricci tensor can be expanded as follows,
\begin{align}
 R_{\mu\nu} \mathbf{d}x^{\mu} \otimes \mathbf{d}x^{\nu} =& - 2 \dot{H} \hat{\bm{e}}^{0} \otimes \hat{\bm{e}}^{0} \nonumber \\
  &+ \left(3 H^2 + \dot{H}\right) \eta_{MN} \hat{\bm{e}}^{M}\otimes\hat{\bm{e}}^{N}\,,
\end{align}
where $\eta_{MN}$ denotes the Minkowski metric tensor with tetrad indices.
Since $H \rightarrow H_{\Lambda}$ and $\dot{H} \rightarrow 0 $ by the condition \eqref{ainf},
the components of the Ricci tensor with respect to the FLRW tetrad are finite.
Thus, there appears to be no ill behavior for comoving timelike observers.
Also, any scalar curvature invariant constructed from the Ricci tensor is finite in the limit to the boundary $\mathscr{B}^{-}$.
Nevertheless, any component of the Ricci tensor with respect to the p.\,p.\ tetrad $\bm{e}^M$ could possibly diverge,
because the p.\,p.\ tetrad is related to the FLRW tetrad through an infinite boost.
Concretely, by using the expressions \eqref{FLRWtoPP}, we can write
\begin{widetext}
\begin{align}
 R_{\mu\nu}\mathbf{d}x^{\mu} \otimes \mathbf{d}x^{\nu} =&~\frac{\dot{H} a_0 ^2}{2 a^2} \left[-\left(1 + \frac{a^2}{a_0^2}\right)^2 \bm{e}^{0}
 \otimes \bm{e}^{0} + 2 \left(1 - \frac{a^4}{a_0^4}\right) \bm{e}^{(0} \otimes \bm{e}^{1)} - \left(1 - \frac{a^2}{a_0^2}\right)^2 \bm{e}^{1} \otimes
 \bm{e}^{1}\right] \nonumber \\
 & + \left(3 H^2 + \dot{H}\right) \eta_{MN} \bm{e}^{M} \otimes \bm{e}^{N} \, ,
\end{align}
\end{widetext}
where in general $\bm{e}^{(M}\otimes\bm{e}^{N)}$ is shorthand notation for
$(\bm{e}^M\otimes\bm{e}^N+\bm{e}^N\otimes\bm{e}^M)/2$.
From the above, we can see that the $(0,0)$, $(0,1)$, and $(1,1)$ components include the possibly divergent quantity $\dot{H}/a^2$
as $t\rightarrow -\infty$ and $a\rightarrow 0$.
Therefore, we arrive at the following statement, the key result of the present work:
\begin{thm}
The boundary $\mathscr{B}^-$ is a p.\,p.\ curvature singularity, or more precisely an intermediate singularity, if
\begin{equation}
 \bigg|\lim_{t\rightarrow -\infty}\frac{\dot H}{a^2}\bigg|=\infty\,, \label{eq:C0inextendibility}
\end{equation}
and consequently, the corresponding inflationary spacetime cannot be extended beyond $\mathscr{B}^{-}$.
On the other hand, if $\dot{H}/a^2$ converges, then the past boundary $\mathscr{B}^{-}$ is not singular,
and it can be extendible at least locally.
\end{thm}

The above statement relies solely on the behavior of the Ricci tensor (not on its derivatives),
so the statement is pertaining to continuous ($C^0$) extendibility and $C^0$ p.\,p.\ curvature singularities.
The precise definition of $C^r$ p.\,p.\ curvature singularities and $C^r$ extendibility for any integer $r\geq 0$
can be found in appendix \ref{Crextendibility}, and as expected, the criterion depends on the behavior of the $r$-th covariant
derivative of the Ricci tensor. For example, continuously differentiable ($C^1$) extendibility is possible if
the first covariant derivative of the Ricci tensor is continuous and does not diverge, i.e., $R_{\mu\nu}$ must be of class $C^1$.
Similarly, smooth (infinitely differentiable or $C^\infty$) extendibility requires the Ricci tensor to be smooth as well.

In what follows, we focus on continuous ($C^0$) extendibility, and that is what is implicitly meant unless specified.
Parallely propagated curvature singularities of class $C^1$, $C^2$ and all the way to $C^\infty$ are `weaker' or `milder' in
the sense that they may only involve the divergence of quantities with higher derivatives of the Hubble parameter (see appendix \ref{Crextendibility}),
e.g.~quantities like $\ddot H/a^3$, $\dddot H/a^4$, etc.

\section{Coordinates beyond the past boundary}
\label{sec:coordinate}

In previous section, we found that an inflationary spacetime is possibly past extendible if the quantity $\dot{H}/a^2$ is finite all the way to
the infinite past. The question that is raised in this case is how one can extend the spacetime beyond the boundary.
In this section, we construct a new set of coordinates for the flat FLRW spacetime
analogously to Eddington-Finkelstein coordinates in Schwarzschild spacetime.
Let us consider a new set of coordinates $\{\lambda, v, \theta, \phi\}$ defined by
\begin{align}
\label{eq:lambdanewcoorddef}
 &\lambda \equiv \lambda[t,- \infty] = \int_{-\infty}^t \mathrm{d}t' \, a(t') \, , \\
 &v \equiv \eta + r\,.
\end{align}
Note that, with these coordinates, the null geodesic \eqref{nullgeotilde} corresponds to the curve characterized by
$v=\theta=\phi=\text{constant}$. Now, the affine parameter is chosen so that $\lambda = 0$ corresponds to the past boundary $\mathscr{B}^{-}$.
Using the relations
\begin{align}
 &\mathrm{d} t = \frac{1}{a}\,\mathrm{d}\lambda\,,\\
 &\mathrm{d} r = \mathrm{d} v - \mathrm{d} \eta = \mathrm{d} v - \frac{1}{a^2} \, \mathrm{d}\lambda \, ,
\end{align}
we can write the FLRW metric in terms of the new coordinates as
\begin{equation}
 g_{\mu\nu}\mathrm{d}x^{\mu}\mathrm{d}x^{\nu} = - 2\,\mathrm{d}\lambda\,\mathrm{d}v + a^2\,\mathrm{d}v^2 +a^2 r^2\,\mathrm{d}\Omega_{(2)}^2\,.\label{glambda}
\end{equation} 
In the limit toward $\mathscr{B}^{-}$, the quantity $a\,r$ converges under the assumption \eqref{ainf}, because it can be evaluated as
\begin{equation}
  a\, r \sim - a\, \eta \sim -\left(\bar{a}\, e^{H_{\Lambda}t}\right)\left(- \frac{1}{\bar{a}\, H_{\Lambda}} e^{-H_{\Lambda} t}\right) = \frac{1}{H_{\Lambda}}\,.
\end{equation}
Thus, the metric is regular at $\mathscr{B}^{-}$, and it is given by
\begin{equation}
 \left.g_{\mu\nu}\mathrm{d}x^{\mu}\mathrm{d}x^{\nu}\right|_{\mathscr{B}^{-}} = - 2\,\mathrm{d}\lambda\,\mathrm{d}v
  +\frac{1}{H_{\Lambda}^2}\,\mathrm{d}\Omega_{(2)}^2\,.\label{gmnB}
\end{equation}
We note that the metric components are regular on $\mathscr{B}^{-}$ even if $\mathscr{B}^-$ is a p.\,p.\ curvature singularity,
because $\dot{H}/ a^2 $ does not appear in the expression \eqref{gmnB}. In the inextendible case, an ill behavior only appears in the
components of the Ricci tensor, which can be expressed as
\begin{equation}
 R_{\mu\nu}\mathrm{d}x^{\mu}\mathrm{d}x^{\nu} = -2 \frac{\dot{H}}{a^2}\,\mathrm{d}\lambda^2
  + \left(3 H^2 + \dot{H}\right) g_{\alpha\beta}\mathrm{d}x^{\alpha}\mathrm{d}x^{\beta}\,.\label{Riccilambda}
\end{equation}
As it is clear from Eqs.~\eqref{gmnB} and \eqref{Riccilambda}, when $\dot{H}/a^2$ converges, there is no ill behavior at $\lambda = 0$.
Thus, the spacetime can be extended to $\lambda\leq 0$ by making use of this coordinate system.

\section{Examples}
\label{sec:examples}

\subsection{Exact de Sitter spacetime}
As a first example of our general discussion, we first demonstrate
the maximal extension of flat de Sitter space, where the scale factor is exactly given by
\begin{equation}
 a(t) = \bar{a}\, e^{H_{\Lambda} t}\,.
\end{equation}
Here, the FLRW time coordinate $t$ is defined in the region $ t \in (-\infty, \infty)$.
Since the affine parameter $\lambda$ can be evaluated as
\begin{equation}
 \lambda = \frac{\bar{a}\, e^{H_{\Lambda} t}}{H_\Lambda}\,,\label{lambdadS}
\end{equation}
the coordinate region $t \in (-\infty, \infty)$ corresponds to $\lambda \in (0 , \infty)$, hence the null geodesics are past incomplete.
From Eq.~\eqref{lambdadS}, the scale factor can be written as a function of $\lambda$ as
\begin{equation}
 a(\lambda) = H_{\Lambda} \lambda\,.
\label{eq:alambdadeSitter}
\end{equation}
Also, the conformal time can be evaluated as
\begin{equation}
 \eta = - \frac{e^{- H_{\Lambda}t}}{\bar{a} H_{\Lambda}} = - \frac{1}{H_{\Lambda}^2 \lambda }\,.\label{etadS}
\end{equation}
Since the coordinate region $t \in (-\infty, \infty)$ corresponds to $\eta \in (- \infty, 0)$, flat de Sitter space is
conformally isometric to the lower half of Minkowski spacetime as shown by the upper triangle of Fig.~\ref{fdS}.
Using Eqs.~\eqref{lambdadS} and \eqref{etadS}, we can write the metric in the coordinates $\{\lambda, v, \theta, \phi\}$ as
\begin{align}
 g_{\mu\nu}\mathrm{d}x^{\mu}\mathrm{d}x^{\nu} = &-2\,\mathrm{d}\lambda\,\mathrm{d}v+H_{\Lambda}^2\lambda^2\,\mathrm{d}v^2 \nonumber \\
  &+ \frac{1}{H_{\Lambda}^2} \left( 1 + H_{\Lambda}^2 \lambda v\right)^2 \mathrm{d} \Omega_{(2)}^2\,.
\end{align}
Now, the metric tensor is well defined even for nonpositive values of $\lambda$.
Thus, we can extend\footnote{In the language of appendix \ref{Crextendibility},
exact de Sitter space is one of the examples where the spacetime is actually smoothly ($C^\infty$) extendible.}
flat de Sitter space with $\lambda \in (0, \infty)$ to the entire de Sitter spacetime with $\lambda \in (- \infty, \infty)$.
Since $\lambda$ is nothing but the affine parameter of a null geodesic, this geodesic is now complete in the entire de Sitter spacetime.
We note that with the closed FLRW coordinates of the entire de Sitter spacetime, also known as the global coordinates, the line element is given by
\begin{equation}
 g_{\mu\nu}\mathrm{d}x^{\mu} \mathrm{d}x^{\nu}
  = - \mathrm{d} t_g^2 + \frac{\cosh^2 (H_{\Lambda} t_g)}{H_{\Lambda}^2} \left(
  \mathrm{d} \psi^2 + \sin^2 \psi \, \mathrm{d} \Omega_{(2)}^2 \right)\,,
\label{eq:metricdSglobalcoordinates}
\end{equation}
where $t_g$ is the global time coordinate, and $\psi\in(0,\pi)$ is the third angle describing the 3-sphere with line element
$\mathrm{d}\Omega^2_{(3)}=\mathrm{d} \psi^2 + \sin^2 \psi \, \mathrm{d} \Omega_{(2)}^2$.
The above metric can be obtained from our coordinates by the following coordinate transformation:
\begin{align}
 \lambda &=  \frac{1}{H_{\Lambda}^2}\big(\cosh[H_{\Lambda} t_g]\cos\psi + \sinh[H_{\Lambda} t_g] \big)\,;\\
 v &= - \frac{1 - e^{H_{\Lambda} t_g}\tan(\psi/2)}{e^{H_{\Lambda}t_g} + \tan(\psi/2)}\,.
\end{align}

\subsection{Inextendible toy model}

\begin{figure}[pt]
\begin{minipage}{0.9\hsize}
\begin{center}
  \includegraphics[width=0.6\hsize]{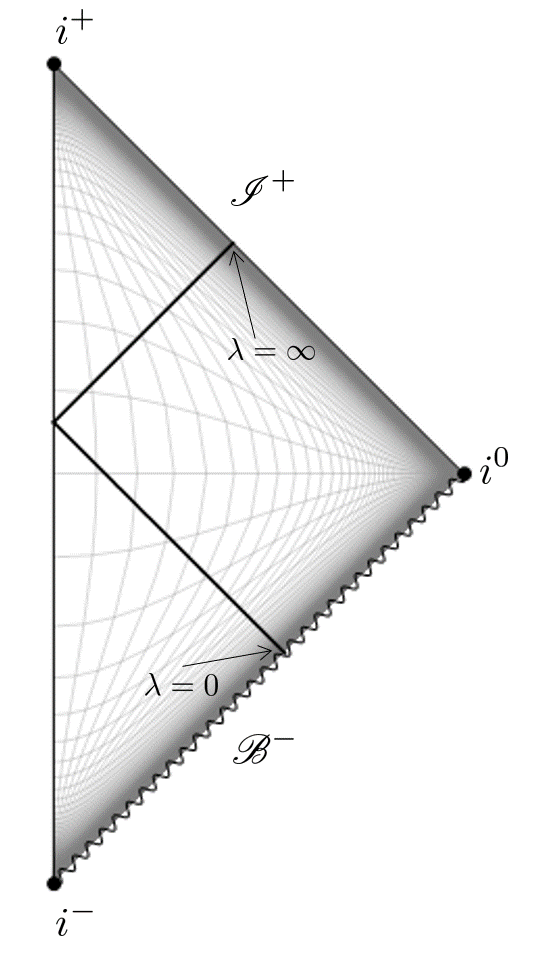}
  \caption{Penrose diagram of the inextendible toy model
  with scale factor given by Eq.~\eqref{amodel1}. The past boundary $\mathscr{B}^{-}$ is a p.\,p.\ curvature
  singularity, and consequently, the spacetime cannot be extended beyond $\mathscr{B}^{-}$.
  The bold line represents a null geodesic, which `starts' on $\mathscr{B}^{-}$ at $\lambda\rightarrow 0$ and `ends' on
  $\mathscr{I}^+$ at $\lambda\rightarrow\infty$.
  Note also that $i^+$ denotes future timelike infinity.}
\label{PDmodel1}
\end{center}
\end{minipage}
\end{figure}

As a toy model, let us consider a flat FLRW spacetime with scale factor given by
\begin{equation}
 a(t) = \frac{\bar{a}\, e^{H_{\Lambda}t}}{1 + \bar{a}\, e^{H_{\Lambda}t}}  
 \simeq
 \begin{cases}
  \bar{a}\, e^{H_{\Lambda}t }  & \text{as}\ t \rightarrow - \infty\,, \\
  1 & \text{as}\  t \rightarrow \infty\,.
 \end{cases}
 \label{amodel1}
\end{equation}
This scale factor represents a universe which starts from de Sitter and approaches Minkowski at $t \rightarrow \infty$.
One can evaluate the conformal time as
\begin{equation}
 \eta = t - \frac{e^{- H_{\Lambda}t}}{\bar{a} H_{\Lambda}}\,,
\end{equation}
and since the coordinate region $t \in (- \infty, \infty)$ corresponds to $\eta \in (- \infty, \infty)$,
this spacetime is conformally isometric to the whole Minkowski spacetime. However, since
\begin{equation}
 \frac{\dot{H}}{a^2} = - \frac{H_\Lambda^2}{\bar{a}\, e^{H_{\Lambda}t}} \rightarrow - \infty\qquad  \text{as}\ t \rightarrow - \infty\,,
\end{equation}
the past boundary $\mathscr{B}^{-}$ is singular (i.e.~it is a p.\,p.\ curvature singularity) according to the key result of the previous section.
The Penrose diagram of this spacetime is shown in Fig.~\ref{PDmodel1}.

Let us explicitly see the inextendible nature of the spacetime by deriving the metric components
in our new coordinate system $\{\lambda, v, \theta, \phi \}$.
Following the definition in Eq.~\eqref{eq:lambdanewcoorddef}, $\lambda$ can be written as
\begin{equation}
 \lambda = H_\Lambda^{-1} \ln\left(1 + \bar{a}\, e^{H_{\Lambda} t} \right)\,.
\end{equation}
Therefore, the scale factor $a$ and the conformal time $\eta$ can be written in terms of $\lambda$ as
\begin{align}
 a(\lambda) &= 1-e^{-H_{\Lambda} \lambda}\,,\\
 \eta(\lambda) & = \frac{1}{H_\Lambda}\left(\ln\left[\frac{e^{H_{\Lambda}\lambda}-1}{\bar{a}}\right]-\frac{1}{e^{H_\Lambda\lambda}-1}\right)\,.
\end{align}
Thus, we can evaluate $a\, \eta$ as
\begin{equation}
 a\, \eta = \frac{e^{-H_{\Lambda}\lambda}}{H_\Lambda}
 \left( \epsilon  \ln \left[\frac{\epsilon}{\bar{a}} \right]-1\right)\,,
\end{equation}
where $\epsilon$ is a function of $\lambda$ defined by 
\begin{equation}
 \epsilon(\lambda) \equiv e^{H_{\Lambda} \lambda} - 1\,.
\end{equation}
Then, the metric can be written as
\begin{align}
 g_{\mu\nu}\mathrm{d}x^{\mu}\mathrm{d}x^{\nu}=&-2\,\mathrm{d}\lambda\,\mathrm{d}v+\left(1-e^{-H_{\Lambda}\lambda}\right)^2\mathrm{d}v^2 \nonumber \\
  &+\frac{e^{-2H_{\Lambda}\lambda}}{H_{\Lambda}^2}\left(\epsilon\ln\left[\frac{\epsilon}{\bar{a}}\right]-1
  -H_{\Lambda}\epsilon v\right)^2\mathrm{d}\Omega_{(2)}^2\,.
\end{align}
As one takes the limit $\lambda \rightarrow 0$ towards $\mathscr{B}^-$, the above expression becomes exactly equal to Eq.~\eqref{gmnB},
and so the metric components are well defined in that limit.
However, the metric is not differentiable in that limit because of the term of the form $\epsilon \ln \epsilon$.
Thus, the Ricci tensor is not $C^0$, and the spacetime cannot be extended beyond $\lambda = 0$.

\subsection{Extendible toy model}

\begin{figure}[pt]
\begin{minipage}{0.9\hsize}
\begin{center}
 \includegraphics[width=0.6\hsize]{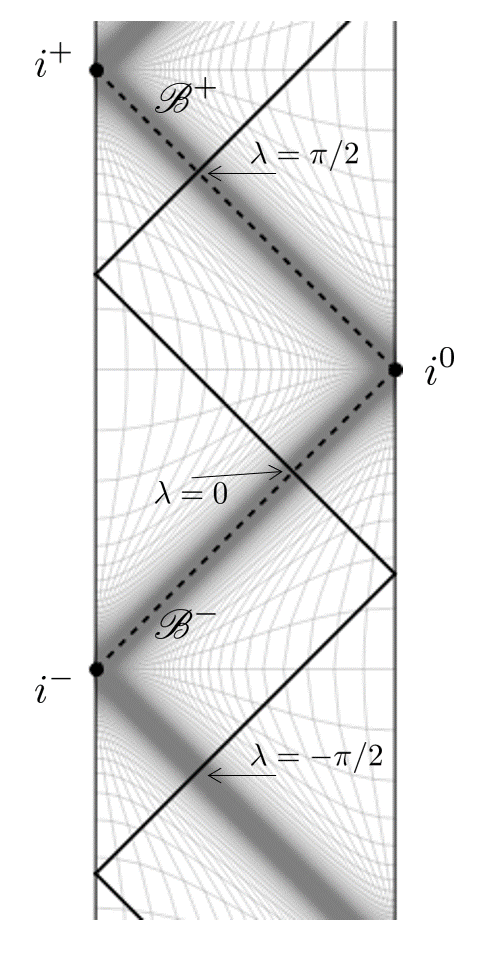}
 \caption{Penrose diagram of the extendible toy model,
  where the scale factor is given by Eq.~\eqref{amodel2}, and we take $\bar{a} =1 $ and $H_{\Lambda} = 1$ (in Planck units) for illustrative purposes.
  The original flat FLRW spacetime corresponds to the triangle region surrounded by dashed lines,
  with $i^{+}$ and $i^{-}$ representing future and past timelike infinity for comoving observers in the original universe.
  The spacetime can be extended beyond the future and past boundaries $\mathscr{B}^{\pm}$, and the whole spacetime is geodesically complete.}
\label{PDmodel2}
\end{center}
\end{minipage}
\end{figure}

Let us consider another toy model with the scale factor given by
\begin{equation}
 a(t) = \frac{\bar{a}}{e^{H_{\Lambda} t}+e^{- H_\Lambda t}}\,.\label{amodel2}
\end{equation}
In the limit $t \rightarrow \pm \infty$, the scale factor behaves as that of a contracting or expanding flat de Sitter universe:
\begin{equation}
 a(t) \simeq
\begin{cases}
 \bar{a}\, e^{  H_{\Lambda} t}& \text{as} \ t\rightarrow - \infty\,; \\
 \bar{a}\, e^{- H_{\Lambda} t}& \text{as} \ t\rightarrow \infty \, .
\end{cases}
\end{equation}
Since the conformal time,
\begin{equation}
 \eta = \frac{e^{H_{\Lambda} t} - e^{- H_{\Lambda}t}}{\bar{a} H_{\Lambda}}\,,
\end{equation}
is defined in the parameter region $\eta \in (-\infty, \infty)$, the flat FLRW region $t \in (-\infty, \infty)$ is again
conformally isometric to the whole Minkowski spacetime. Since $\dot{H}/a^2$ is now finite,
\begin{equation}
 \frac{\dot{H}}{a^2} = -\frac{4 H_{\Lambda}^2}{\bar{a}^2}\,,
\end{equation}
any component of the Ricci tensor is well behaved in the limit to the past null boundary $\mathscr{B}^{-}$.
Therefore, this spacetime is extendible\footnote{The toy model here is actually smoothly ($C^\infty$)
extendible just like exact de Sitter space. This is due to the fact that $\dot H/a^2=\mathrm{constant}$,
i.e.~$\dot H\sim a^2$, and it is shown in appendix \ref{Crextendibility} that $C^\infty$ extendibility follows if $\dot H\sim a^q$
as $a\rightarrow 0$ with $q\in\mathbb{Z}_{\geq 2}$. The present case corresponds to $q=2$.}
beyond the past null boundary the same way flat de Sitter space can be.
The affine parameter $\lambda$ is obtained as follows,
\begin{equation}
 \lambda = \frac{\bar{a}}{H_{\Lambda}}\arctan e^{H_{\Lambda}t}\,,
\end{equation}
and so the region $t \in (-\infty, \infty)$ corresponds to $\lambda \in (0, \bar{a} \pi/ 2H_\Lambda )$.
Hence the original region is both future and past incomplete. Let us denote the future null boundary at $\lambda = \bar{a} \pi /2 H_\Lambda $
by $\mathscr{B}^{+}$. The scale factor can then be written in terms of $\lambda$ as
\begin{equation}
 a(\lambda) = \frac{\bar{a}}{2}\sin\left(\frac{2 H_\Lambda \lambda}{\bar{a}}\right)\,,
\label{eq:alambdaextendibletoymodel}
\end{equation}
hence
\begin{equation}
 a(\lambda) \eta(\lambda) = - \frac{1}{H_{\Lambda}} \cos\left(\frac{2 H_\Lambda \lambda}{\bar{a}}\right),
\end{equation}
and so the metric in the original region $\lambda \in (0, \frac{\bar{a}}{H_\Lambda} \frac{\pi}{2})$ can be written as
\begin{align}
 g_{\mu\nu}\mathrm{d}x^{\mu}\mathrm{d}x^{\nu}=&-2\,\mathrm{d}\lambda\,\mathrm{d}v
 +\frac{\bar{a}^2}{4}\sin^2(2 H_\Lambda\lambda/\bar{a})\,\mathrm{d}v^2 \nonumber \\
  &+\frac{1}{4H_\Lambda^2}\bigg(2\cos\left[\frac{2H_\Lambda\lambda}{\bar a}\right] \nonumber \\
  &\qquad+\bar aH_\Lambda v\sin\left[\frac{2H_\Lambda\lambda}{\bar a}\right]\bigg)^2\mathrm{d}\Omega_{(2)}^2\,.
\label{eq:metricextendibletoymodel}
\end{align}
The important point here is that the scale factor and the components of the metric are well defined in the whole
parameter region $\lambda \in (- \infty, \infty)$, and therefore, we can extend the spacetime to that entire region.
Since the scale factor is periodic, each spacetime region is characterized by the quantity
\begin{equation}
 \frac{2}{\pi}\frac{H_{\Lambda}}{\bar{a}}\lambda \in \left(n,n+1\right)\,,
\end{equation}
where $n\in\mathbb{Z}$ denotes the $n$-th copy of the original spacetime.
The Penrose diagram of the entire spacetime is depicted in Fig.~\ref{PDmodel2}.

The resulting maximally extended geodesically complete spacetime is a cyclic universe with periodically repeating phases of expansion and contraction,
and the transitions from contraction to expansion at $\lambda=n\pi\bar a/2H_\Lambda$ are bounces.
With $a(\lambda)$ given by Eq.~\eqref{eq:alambdaextendibletoymodel}, we notice that $a(\lambda=n\pi\bar a/2H_\Lambda)=0$,
so the scale factor vanishes at each bounce point. Yet, those bounces are nonsingular
since the boundary region at $\lambda=n\pi\bar a/2H_\Lambda$ cannot be described by the usual FLRW coordinate system.
Rather, we see that the metric of Eq.~\eqref{eq:metricextendibletoymodel} reduces
to Eq.~\eqref{gmnB} when $\lambda=n\pi\bar a/2H_\Lambda$, which is the correct description of the nonsingular bouncing surfaces.
In other words, the `physical' scale factor is the one with respect to the comoving observer in the flat FLRW spacetime, and it is
meaningless on $\mathscr{B}^{\pm}$ (or on any other boundary region at $\lambda=n\pi\bar a/2H_\Lambda$).
This is very similar to the case of de Sitter space, where at $\lambda = 0$, the scalar factor given by Eq.~\eqref{eq:alambdadeSitter}
would appear to be singular. However, there exists a coordinate system in which the bounce is explicitly nonsingular,
as can be seen from Eq.~\eqref{eq:metricdSglobalcoordinates}.

\section{Implications for inflationary models}
\label{inflationarymodel}

\subsection{Energy condition for subleading component}

We have seen that convergence of the quantity $\dot{H}/a^2$ is crucial to extend
an inflationary cosmology beyond the past null boundary $\mathscr{B}^{-}$.
Since there is no contribution from vacuum energy to $\dot{H}$ in Einstein gravity,
the behavior of $\dot{H}/a^2$ is determined by the next-to-leading order contribution to the energy density during inflation.
In order to describe this situation effectively, let us consider Einstein gravity with a cosmological constant
and a fluid which follows the equation of state, 
\begin{equation}
 P = w \rho\,,
\end{equation}
where $w$ is a constant, $P$ and $\rho$ are the pressure and the energy density of the fluid, respectively.
We note, though, that a fluid description with the above equation of state might not be a fully realistic situation for inflation,
but it serves as an effective description.
By solving the conservation equation for the fluid as usual, we obtain
\begin{equation}
 \rho \propto a^{- 3 (1 + w)}\,.
\end{equation}
We now assume that $w < -1$ so that the vacuum energy is dominant in the total energy density, i.e.
\begin{equation}
 \rho_\mathrm{tot} = \Lambda + \rho \simeq \Lambda\qquad\mathrm{as}\ a\rightarrow 0\,.
\end{equation}
In that case, the evolution of the Hubble parameter is dominated by the vacuum energy $\Lambda$,
and our assumption \eqref{ainf} is realized.
Then using the Friedman equations, we can write the key quantity $\dot{H}/a^2$ as a function of the scale factor $a$, 
\begin{equation}
 \frac{\dot{H}}{a^2} = - \frac{1}{2 M_{\mathrm{Pl}}^2} \frac{\rho + P}{a^2} \propto \frac{1 + w}{a^{5+3 w}}\,,
\end{equation}
which converges in the limit $a \rightarrow 0$ only if $w = -1$ or $5+ 3 w \leq 0$.
The former case exactly corresponds to the vacuum energy $\Lambda$.
Thus, if there is a correction to the vacuum energy, the equation of state of the next-to-leading order component has to satisfy the latter condition,
\begin{equation}
 w \leq - \frac{5}{3}\,.
\end{equation}
This clarifies the discontinuous nature between exact de Sitter space and inflationary cosmology satisfying Eq.~\eqref{ainf}:
although the spacetime is nonsingular if $w$ is exactly equal to $-1$, any arbitrary small deviation from $w = -1$ leads to a
p.\,p.\ curvature singularity.
The singularity is avoided only if the next-to-leading order component to the energy density
goes to zero fast enough as $a\rightarrow 0$.
The above result shows that it is not enough to violate the Null Energy Condition with $w<-1$
since an equation of state parameter in the range $w\in(-5/3,-1)$ would still lead to a p.\,p.\ curvature singularity.

We note that the condition $w\leq -5/3$ generally ensures $C^0$ extendibility. However, if $-5-3w$ is an integer in addition to
the condition $w\leq -5/3$ (i.e.~if $w\in\mathbb{Z}_{\leq -5}/3$), then the spacetime is $C^\infty$ extendible
as shown in appendix \ref{Crextendibility}. If $w$ is not an integer multiple of $1/3$ but still $w\leq -5/3$, then the spacetime is at most
$C^{\lfloor-3w\rfloor-5}$ extendible.

\subsection{Single field slow-roll inflation}

Let us derive the expression for the key quantity $\dot{H}/a^2$ in the case of slow-roll inflation models
driven by a canonical scalar field $\varphi$ with a potential $V(\varphi)$.
The complete set of equations of motion is given by\footnote{From here on, a prime denotes a derivative with respect to the argument of the function.}
\begin{equation}
 H \simeq \sqrt{\frac{V(\varphi)}{3 M_{\mathrm{Pl}}^2}}\,,\qquad \dot{\varphi} \simeq - \frac{V'(\varphi)}{3 H}\,,\label{slowrolleom}
\end{equation}
provided the following slow-roll approximations are satisfied:
\begin{equation}
\label{eq:slowrollapprox}
  \frac{|\ddot{\varphi}|}{3H|\dot{\varphi}|} \ll 1\,,\qquad  \frac{\dot{\varphi}^2}{2 V(\varphi)} \ll 1\,.
\end{equation}
From these equations, we can write the scale factor as a function of $\varphi$, 
\begin{equation}
 a(\varphi) = a_e e^{-\mathcal{N}(\varphi)}\,,\label{slowrolla}
\end{equation}
with $\mathcal{N}(\varphi)$, the e-folding number, given by
\begin{equation}
\label{eq:Ne-fold}
 \mathcal{N}(\varphi) \simeq \frac{1}{M_{\mathrm{Pl}}^2} \int^{\varphi}_{\varphi_e} \mathrm{d}\varphi~\frac{V}{V'} \,.
\end{equation}
Here, $a_e$ and $\varphi_e$ are the values of $a$ and $\varphi$ at the end of inflation respectively.

Using Eqs.~\eqref{slowrolleom} and \eqref{slowrolla}, we can then write the key quantity $\dot{H}/a^2$ in terms of $\varphi$,
\begin{equation}
 \frac{\dot{H}}{a^2}\simeq-\frac{1}{6a^2}\frac{(V')^2}{V}=-\frac{1}{6 a_e^2}\frac{(V')^2}{V} e^{2 \mathcal{N}(\varphi)}
  =-\frac{1}{6 a_e^2} f(\varphi)\,,
\end{equation}
where the function $f$ is defined by
\begin{equation}
 f(\varphi)\equiv\frac{(V')^2}{V}e^{2\mathcal{N}(\varphi)}\simeq\frac{(V')^2}{V}\exp\left(\frac{2}{M_{\mathrm{Pl}}^2}
  \int^{\varphi}_{\varphi_e}\mathrm{d}\varphi~\frac{V}{V'}\right)\,.\label{deff}
\end{equation}
Thus, a necessary condition for a slow-rolling inflationary cosmology to be free of p.\,p.\ singularities can be written as
\begin{equation}
 \lim_{\varphi \rightarrow \varphi(-\infty)} f(\varphi) = \mathrm{finite} \,, \label{limf}
\end{equation}
where $\varphi(-\infty)$ is the value of $\varphi(t)$ in the limit $t \rightarrow - \infty$.
Thus, for any given potential $V(\varphi)$, we can judge the presence of a singularity by evaluating \eqref{deff} and its limit.
In the rest of this section, we evaluate \eqref{limf} for the Starobinsky model and for a small field inflationary model.

\subsubsection*{Starobinsky model}

Let us consider the Starobinsky model \cite{Starobinsky:1980te} with Einstein frame potential given by
\begin{equation}
 V(\varphi) = \frac{3}{4} m^2 M_{\mathrm{Pl}}^2 \left(1 - e^{-\sqrt{\frac{2}{3}} \frac{\varphi}{M_{\mathrm{Pl}}}}\right)^2\,,
\end{equation}
where $m$ is the `mass' of the inflaton.
Inflation occurs at large positive field values,
and $\varphi$ slowly rolls toward $0$ as time $t$ increases.
Inversely, $\varphi$ approaches $+\infty$ in the limit $t \rightarrow -\infty$.
In that limit, $V(\varphi)\simeq 3 m^2 M_\mathrm{Pl}^2 / 4=\mathrm{constant}$,
so the potential acts like a cosmological constant, i.e.~the spacetime is asymptotically de Sitter.
With the above potential, the e-folding number $\mathcal{N}$ can be evaluated following Eq.~\eqref{eq:Ne-fold}, and one finds
\begin{equation}
 \mathcal{N}(\varphi) \simeq \frac{3}{4}e^{\sqrt{\frac{2}{3}}\frac{\varphi}{M_{\mathrm{Pl}}} } \qquad \mathrm{as}\ \varphi \rightarrow \infty\,.
\end{equation}
Since the quantity $(V')^2/V$ is given by
\begin{equation}
\label{eq:Vp2VS}
 \frac{(V')^2}{V} = 2 m^2 e^{-2 \sqrt{\frac{2}{3}}\frac{\varphi}{M_{\mathrm{Pl}}}}\,,
\end{equation}
it cannot suppress the factor of $e^{2\mathcal{N}}$ in the expression for $f(\varphi)$. Indeed,
\begin{align}
 f(\varphi) &= 2 m^2  e^{ 2 \mathcal{N}(\varphi)- 2 \sqrt{\frac{2}{3}}\frac{\varphi}{M_{\mathrm{Pl}}}} \nonumber \\
 &\simeq 2 m^2  \exp\left( \frac{3}{2} \exp\left[\sqrt{\frac{2}{3}} \frac{\varphi}{M_{\mathrm{Pl}}}\right]\right) \rightarrow \infty
\end{align}
as $\varphi\rightarrow\infty$. We see from Eq.~\eqref{eq:Vp2VS}, which is proportional to $\dot H$,
that the Hubble parameter approaches a constant exponentially fast
as $\varphi\rightarrow\infty$ in field space.
This matches the intuition that Starobinsky inflation rapidly approaches de Sitter in field space (at large field values).
However, the scale factor reaches zero even faster than $\dot H$ as $a\sim\exp(-\exp(\varphi))$.
The subtlety comes from the fact that the potential is very flat at large field values,
which implies that large time intervals are needed for small field displacements.
Consequently, de Sitter is actually approached only very slowly in physical time compared to the rate at which the scale factor goes to zero.
Thus, the ratio $\dot H/a^2$ and equivalently $f(\varphi)$ blow up, and the spacetime is inextendible.
We can conclude that if Starobinsky inflation starts from the infinite past
at $t \rightarrow -\infty$ for comoving observers, then the past boundary $\mathscr{B}^{-}$ must be a p.\,p.\ curvature singularity.

We note, however, that Starobinsky inflation is unlikely to start from the infinite past in the first place. Indeed, this would require
the initial field velocity to exactly vanish, which represents extreme fine-tuning.
In general, the field equation of motion is $\ddot\varphi+3H\dot\varphi\simeq 0$ for a nearly flat potential,
and with $\dot\varphi\neq 0$ initially, this implies $\dot\varphi\propto a^{-3}$ and $\rho\sim a^{-6}$ (kinetic domination as $a\rightarrow 0$).
Accordingly, the first slow-roll approximation in Eq.~\eqref{eq:slowrollapprox} would not be satisfied.
This is known as ultra-slow-roll or non-attractor inflation (see, e.g., \cite{Tsamis:2003px,Kinney:2005vj,Cai:2016ngx,Dimopoulos:2017ged}).
In that situation, the effective equation of state parameter $w$ would tend to unity as $a\rightarrow 0$,
and the Universe would be past incomplete (standard Big Bang curvature singularity).

Another comment is in order: Starobinsky inflation with the Einstein frame potential $V(\varphi)$ given above
is equivalent to an $f(R)$ modified theory of gravity of the form $f(R)=R+R^2/(6m^2)$ in the Jordan frame after a conformal transformation.
It would be natural to extend the above analysis to inflationary scenarios with different $f(R)$ theories of gravity,
e.g., slight deviations from Starobinsky inflation or generalizations thereof (see, e.g., \cite{Miranda:2017juz}).
This shall be the subject of a follow-up study.

\subsubsection*{Small Field inflation}

Let us consider another slow-roll inflation model with a Higgs-like potential \cite{Vilenkin:1994pv,Linde:1994wt} of the form
\begin{equation}
 V(\varphi) = V_0 \left(1 - \left(\frac{\varphi}{2 m}\right)^2 \right)^2\,,
\end{equation}
where $V_0$ is a positive constant, and $m$ is another mass scale.
We would like to focus on small field inflation, which occurs when\footnote{Chaotic inflation is also possible when $\varphi$ is much larger than $m$,
but we note that in the case of such chaotic inflation models
(and more generally for $V(\varphi)\propto\varphi^p$, $p>0$), the inflaton potential
energy diverges as $\varphi \rightarrow \infty$ (i.e.~$t\rightarrow - \infty$).
Consequently, the effective description of the inflationary cosmology based on classical gravity would
no longer be valid near $\mathscr{B}^{-}$. Thus, the present analysis cannot address the past extendibility of such
inflationary models.} $\varphi \ll m$.
We note, however, that such small field inflation models
are unstable against initial condition fluctuations \cite{Goldwirth:1991rj,Brandenberger:2016uzh}.
From the above potential, we find that the e-folding number is given by
\begin{equation}
 \mathcal{N}(\varphi)=-\frac{m^2}{M_{\mathrm{Pl}}^2}\ln\left(\frac{\varphi}{\varphi_e}\right) + \frac{\varphi^2 - \varphi_e^2}{8 M_{\mathrm{Pl}}^2}\,,
\end{equation}
and it diverges in the limit $\varphi \rightarrow 0$.
Thus, the limit $t \rightarrow -\infty$ corresponds to the limit $\varphi \rightarrow 0$. Then, $f(\varphi)$ can be evaluated as follows,
\begin{equation}
 f(\varphi) = \frac{V_0}{m^4} \varphi^2 \left(\frac{\varphi}{\varphi_e}\right)^{- 2 \frac{m^2}{M_{\mathrm{Pl}}^2}}
 e^{\frac{\varphi^2 - \varphi_e^2}{4 M_{\mathrm{Pl}}^2}}\,,
\end{equation}
and one finds that $f(\varphi)$ converges in the limit $\varphi \rightarrow 0$ when $m \leq M_{\mathrm{Pl}}$.
Thus, if small field inflation starts from the infinite past at $t \rightarrow -\infty$ for comoving observers, then the
Universe can be continuously ($C^0$) extended beyond the past boundary $\mathscr{B}^{-}$. In other words,
noncomoving geodesics exit the original inflationary region sufficiently far in the past.
However, the above does not tell us whether the past boundary $\mathscr{B}^{-}$ is $C^r$ extendible for $r\geq 1$,
and the spacetime could very well be $C^1$ inextendible. This remains to be verified, but the condition \eqref{limf}
would be much more complicated.

\subsection{Limiting curvature models}

In this subsection, we demonstrate how to evaluate the key quantity $\dot H/a^2$ in the limit corresponding to de Sitter in a class
of modified gravity models. Specifically, we focus on a gravitational theory with limiting curvature,
which is claimed to have nonsingular cosmological solutions,
as proposed and investigated in Refs.~\cite{Mukhanov:1991zn,Brandenberger:1993ef,Yoshida:2017swb}
(see also references therein, and Refs.~\cite{Easson:1999xw,Easson:2006jd,Chamseddine:2016uef}
and Refs.~\cite{Trodden:1993dm,Chamseddine:2016ktu,Yoshida:2018kwy} for other applications to cosmological and black hole spacetimes respectively).
Let us briefly review the theory and the inflationary solutions investigated in Ref.~\cite{Yoshida:2017swb}. The action of this theory is given by
\begin{equation}
\label{eq:actionlimitingcurvature}
 S = \frac{M_{\mathrm{Pl}}^2}{2} \int \mathrm{d}^4 x\,\sqrt{-g} \left( R + \sum_{j=1}^2\big[\chi_j I_j - V_j(\chi_j)\big]\right),
\end{equation}
where $\chi_1$ and $\chi_2$ are scalar fields, and $I_1$ and $I_2$ are curvature invariant functions, which reduce to
\begin{equation}
 I_1^\mathrm{FLRW} = 12 H^2\,,\qquad   I_2^\mathrm{FLRW} = - 6 \dot{H}\,, \label{I12}
\end{equation}
in a flat FLRW spacetime. We note that there are many choices of curvature invariant functions which satisfy the condition \eqref{I12},
and the stability of the cosmological perturbations\footnote{See Ref.~\cite{Yoshida:2017swb} for examples of
covariant curvature invariant functions that reduce to Eq.~\eqref{I12}
and for the analysis of the perturbations.} strongly depends on the choice of $I_1$ and $I_2$.
However, the background dynamics can be uniquely determined only from the condition \eqref{I12}.

Varying the action \eqref{eq:actionlimitingcurvature} with respect to $\chi_1$ and $\chi_2$ gives rise to the following constraint equations
at the background level,
\begin{equation}
 12H^2=~V_1'(\chi_1)\,,\qquad   -6\dot H=~V_2'(\chi_2)\,,  \label{eq:dotH} 
\end{equation}
which ensure the finiteness of $H$ and $\dot{H}$ if $V_1'$ and $V_2'$ are finite for any value of $\chi_1$ and $\chi_2$.
This is the mechanism to limit the divergence of the curvature invariants in this theory.
The other independent equation of motion is given by (see Ref.~\cite{Yoshida:2017swb} for a derivation)
\begin{equation}
 ~(1-2\chi_1-3\chi_2)H^2-\dot{\chi}_2H-\frac{V_1+V_2}{6} = 0\,.
\end{equation}
The above equations of motion can be rewritten as a set of first-order differential equations only involving $a$, $\chi_1$, and $\chi_2$.
In particular, the $\chi_1-\chi_2$ phase space trajectories are governed by the following equation:
\begin{equation}
 \frac{\mathrm{d}\chi_2}{\mathrm{d}\chi_1}=\frac{V_1''}{4V_2'}\left(3\chi_2+2\chi_1-1+\frac{2(V_1+V_2)}{V_1'}\right)\,.
\label{eq:dx2dx1}
\end{equation}
Similarly, the solutions in the $\chi_2 - a$ space satisfy the following equation:
\begin{equation}
\label{eq:dx2da}
 \frac{\mathrm{d}\chi_2}{\mathrm{d}\ln a}=-\left(3\chi_2+2\chi_1-1+\frac{2(V_1+V_2)}{V_1'}\right)\,.
\end{equation}

\begin{figure}[pt]
\begin{minipage}{0.9\hsize}
\begin{center}
 \includegraphics[width=0.9\hsize]{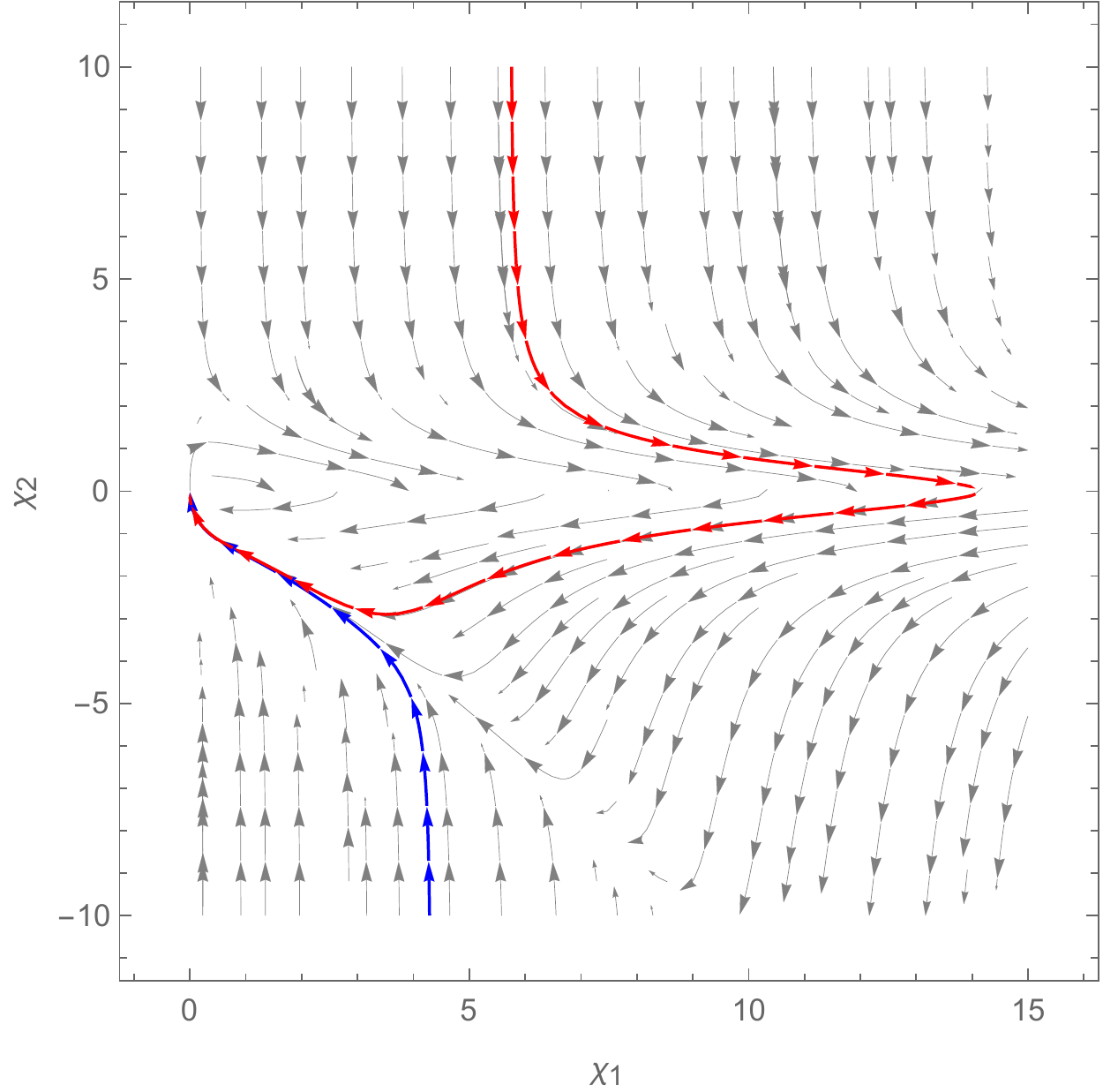}
 \caption{Trajectories in the $\chi_1-\chi_2$ phase space following Eq.~\eqref{eq:dx2dx1} with the potentials \eqref{eq:V1} and \eqref{eq:V2}.
 The arrows point forward in time. In the limit $t \rightarrow - \infty$, there are two kinds of trajectories:
 those that go to $\chi_1 \rightarrow \text{constant}$ and $\chi_2 \rightarrow +\infty$ (an example is highlighted in red)
 and those that go to $\chi_1 \rightarrow \text{constant}$ and $\chi_2 \rightarrow - \infty$ (an example is highlighted in blue).}
\label{trajectory}
\end{center}
\end{minipage}
\end{figure}

In order to determine if an inflationary solution [one which satisfies Eq.~\eqref{ainf}] is past (in)complete,
we need to check the limit of the ratio
$\dot H/a^2$ as $a\rightarrow 0$ (which is equivalent to the limit $t\rightarrow -\infty$ when Eq.~\eqref{ainf} is satisfied).
In order for the spacetime to be asymptotically de Sitter, the potentials are going to be chosen as follows \cite{Brandenberger:1993ef,Yoshida:2017swb}:
\begin{align}
\label{eq:V1}
 V_1(\chi_1)&=12H_\mathrm{max}^2\frac{\chi_1^2}{1+\chi_1}\left(1-\frac{\ln(1+\chi_1)}{1+\chi_1}\right)\,,\\
\label{eq:V2}
 V_2(\chi_2)&=-12H_\mathrm{max}^2\frac{\chi_2^2}{1+\chi_2^2}\,,
\end{align}
where $H_\mathrm{max}$ is a positive constant. Then, one can use Eq.~\eqref{eq:dx2dx1} to draw the phase space trajectories as shown
in Fig.~\ref{trajectory}.
As it is clear from the diagram\footnote{We note that only the region where $\chi_2 < 0$ was plotted in Ref.~\cite{Yoshida:2017swb}.
This region corresponds to the region $\dot{H} < 0$,
because $I_1$ and $I_2$ satisfy the condition \eqref{I12} only when $\dot{H} < 0$ in the model investigated there.
However, in general, some trajectories enter the region where $\chi_2 \geq 0$ if $I_1$ and $I_2$ are appropriately defined in this region.}
in the limit $t \rightarrow - \infty$, trajectories go to $\chi_1 \rightarrow \text{constant}$ and $\chi_2 \rightarrow \pm \infty$,
and in that limit, the trajectories are asymptotically de Sitter \cite{Brandenberger:1993ef}.
Taking the limit $|\chi_2|\rightarrow\infty$ with $\chi_1$ kept constant, Eq.~\eqref{eq:dx2da} with the potentials \eqref{eq:V1} and \eqref{eq:V2}
reduces to
\begin{equation}
 \frac{\mathrm{d}\chi_2}{\mathrm{d}\ln a} \simeq - 3 \chi_2\,,
\end{equation}
and upon integration, the solution is
\begin{equation}
 \chi_2(a)\simeq\frac{K}{a^3}\,,
\end{equation}
where $K$ is an integration constant. Substituting this solution into Eq.~\eqref{eq:dotH}
with the potential of Eq.~\eqref{eq:V2}, one obtains
\begin{equation}
 \frac{\dot H(a)}{a^2} \simeq \frac{1}{a^2} \frac{4 K H_\mathrm{max}^2a^9}{(K^2+a^6)^2} \simeq \frac{4 H_\mathrm{max}^2}{K^3} a^7  \rightarrow 0
 \qquad\mathrm{as}\ a \rightarrow 0\,.
\end{equation}
Therefore, the above trajectories that are asymptotically de Sitter are \emph{not} past incomplete;
one could construct an extension beyond the past boundary $\mathscr{B}^{-}$.
More precisely, the spacetime is smoothly ($C^\infty$) extendible since $\dot H\sim a^q$
with $q=9\in\mathbb{Z}_{\geq 2}$ (see appendix \ref{Crextendibility}).

Finally, we would like to comment on the stability of these solutions.
The stability against cosmological perturbations strongly depends on the choice of
$I_1$ and $I_2$, which satisfy the condition \eqref{I12}.
In Ref.~\cite{Yoshida:2017swb}, the stability for two kinds of curvature invariant functions was investigated.
From those results, it appears the trajectories that are asymptotically de Sitter
are unstable in the infinite past (when $\chi_1 \rightarrow \text{constant}$ and $\chi_2 \rightarrow \infty$).
Therefore, although the solutions shown above are examples of past
extendible inflationary cosmologies in modified gravity, they remain at the level of toy models that cannot describe
the real Universe.

\section{Summary and Discussion}

In the present paper, we showed that an inflationary spacetime with flat spatial curvature [i.e.~a flat FLRW spacetime with a scale factor
which satisfies the condition \eqref{ainf}] has a p.\,p.\ curvature singularity if the quantity $\dot{H}/a^2$ diverges in the limit
$t \rightarrow - \infty$. On the other hand, if $\dot{H}/a^2$ converges, then the past boundary is regular and continuously ($C^0$) extendible.
We presented concrete examples of both inextendible and extendible models in Sec.~\ref{sec:examples}. 
In the context of Einstein gravity with a cosmological constant and a perfect fluid, which follows
the equation of state $p/\rho = w = \text{constant}$, we found that the p.\,p.\ curvature singularity
is only avoidable if $w \leq - 5/3$. In the case of slow-roll inflation
with a canonical scalar field, the key quantity $\dot{H}/a^2$ can be written in terms of the inflaton potential,
and we derived the condition to judge whether the past boundary is singular or not for the given potential.
By using this formula, we found that Starobinsky inflation has a $C^0$ p.\,p.\ curvature singularity,
but a small field inflation model does not. Moreover, in the context of a theory of modified gravity with limiting curvature
as investigated in Refs.~\cite{Brandenberger:1993ef,Yoshida:2017swb},
we computed the asymptotic expression for $\dot{H}/a^2$ and determined that the inflationary solutions are smoothly ($C^\infty$) extendible.

Throughout this paper, we have discussed extendibility of flat, asymptotically de Sitter, FLRW spacetimes.
Of course, it could be possible that there is a noninflationary epoch before inflation or that inflation never occurs in the
very early universe. This would necessarily happen if the vacuum energy became subdominant in the limit $a \rightarrow 0$.
In Einstein gravity, candidates are, for example, spatial curvature with $\rho\propto a^{-2}$,
dust with $\rho\propto a^{-3}$, radiation with $\rho\propto a^{-4}$, or anisotropies with $\rho\propto a^{-6}$.
If there is a positive spatial curvature component,
the early stages of inflation would be described by closed de Sitter space,
where the universe enters a contracting phase sufficiently far in the past.
In the case of a negative spatial curvature, we expect the situation to be similar
to the flat case, because open de Sitter space also has a past boundary $\mathscr{B}^-$ that is extendible.
If some thermal matter components or anisotropies are dominant over other components, it would lead to
a Big Bang initial singularity or a Belinsky-Khalatnikov-Lifshitz singularity \cite{Belinsky:1970ew}.
However, this might not be the case in a quantum theory of gravity (see, e.g., \cite{Lam:2016kmt}).
We stress here that the entire analysis performed in this paper is in the realm of classical General Relativity.
The situation is certainly expected to be different when we better understand the quantum gravity effects at high energies.

We also only considered spacetimes that are perfectly homogeneous and isotropic. Thus, we did not include
the effects of anisotropies or cosmological perturbations. The presence of cosmological perturbations would most likely change
the criterion for past extendibility, as known in the case of eternal inflation models \cite{Linde:1986fd}.
In such a case, our analysis is not applicable. Therefore, an interesting direction is to investigate
how to develop the present analysis in the context of eternal inflation models, as it was done for the singularity theorems
by Borde and Vilenkin \cite{Borde:1993xh,Borde:1996pt}.

\begin{acknowledgments}
We would like to thank Robert Brandenberger, Alan Guth, Ryo Namba, and Ziwei Wang for useful comments.
D.\,Y.~would also like to thank Masahide Yamaguchi, Tsutomu Kobayashi, and Masaru Siino for fruitful discussions.
D.\,Y.~is supported by the Japan Society for the Promotion of Science (JSPS) Postdoctoral Fellowships for Research Abroad.
J.\,Q.~acknowledges financial support from the Vanier Canada Graduate Scholarship
administered by the Natural Sciences and Engineering Research Council of Canada (NSERC).
\end{acknowledgments}

\appendix

\section{General extendibility}\label{Crextendibility}

In this appendix, let us precisely define the concepts of singularity and extendibility. For fully rigorous mathematical definitions
and theorems, we refer to Refs.~\cite{Ellis:1977pj,Clarke1973,Clarke1982,Hawking:1973uf}.

Let us call a spacetime $(\mathcal{M},\bm{g})$, where $\mathcal{M}$ is the manifold and $\bm{g}$ the metric tensor,
to be of class $C^r$ when $C^r$ ($r\in\mathbb{Z}_{\geq 0}$) is the differentiability class of the Riemann tensor.
This is equivalent to the metric tensor being of class $C^{r+2}$.
Then, a boundary $\mathscr{B}\subset\mathcal{M}$ is a $C^r$ p.\,p.\ curvature singularity if any component
of the $r$-th covariant derivative of the Riemann tensor is not of class $C^0$ in the limit toward $\mathscr{B}$
when measured in a p.\,p.\ tetrad basis $\bm{e}^M$.
In particular, if the spacetime is conformally flat, i.e.~if the Weyl tensor vanishes, then the Riemann tensor
is fully determined by the Ricci tensor. In that case, $\mathscr{B}$ is a $C^r$ p.\,p.\ curvature singularity if the quantity
\begin{equation}
 \nabla_{\mu_1}\nabla_{\mu_2}\cdot\cdot\cdot\nabla_{\mu_r}R_{\mu_{r+1}\mu_{r+2}}\bigotimes_{j=1}^{r+2}\mathbf{d}x^{\mu_j}
\end{equation}
expanded in terms of the $\bm{e}^M$'s diverges in the limit toward $\mathscr{B}$.
Then, we say that the spacetime is $C^r$ extendible if and only if there is no $C^r$ p.\,p.\ curvature singularity.
Similarly, the spacetime is $C^r$ inextendible if and only if there is a $C^r$ p.\,p.\ curvature singularity.

Following the above statements, the requirement derived at the end of Sec.~\ref{sec:pptetrad} for an inflationary spacetime
is pertaining to a $C^0$ p.\,p.\ singularity and $C^0$ (continuous) (in)extendibility.
However, the statement can be generalized.
Starting with one covariant derivative of the Ricci tensor in FLRW, one can write
\begin{align}
 \nabla_\sigma R_{\mu\nu}\mathbf{d}x^\sigma\otimes&~\mathbf{d}x^\mu\otimes\mathbf{d}x^\nu=-2(\ddot H-2\dot HH)(\hat{\bm{e}}^0)^{\otimes 3} \nonumber \\
 &+4\dot HH\eta_{MN}\hat{\bm{e}}^M\otimes\hat{\bm{e}}^{(N}\otimes\hat{\bm{e}}^{0)} \nonumber \\
 &+(\ddot H+6\dot{H}H)\eta_{MN}\hat{\bm{e}}^0\otimes\hat{\bm{e}}^{M}\otimes\hat{\bm{e}}^{N}
\label{eq:nablaRmunupp}
\end{align}
in the FLRW tetrad basis. Thus, when assumption \eqref{ainf} is satisfied, $\ddot H\rightarrow 0$ and $\dot{H}H\rightarrow 0$,
and there appears to be no scalar polynomial curvature singularity. However, when transforming to the p.\,p.\ tetrad basis, one finds
\begin{equation}
 \nabla_\sigma R_{\mu\nu}\mathbf{d}x^\sigma\otimes\mathbf{d}x^\mu\otimes\mathbf{d}x^\nu\simeq-\frac{(\ddot{H}-2\dot HH)a_0^3}{4a^3}(\bm{e}^0-\bm{e}^1)^{\otimes 3}
\end{equation}
to leading order in the limit $a\rightarrow 0$.
Therefore, as the spacetime approaches de Sitter, the Hubble parameter is asymptotically a constant, and one needs to check the convergence
of two quantities, $\ddot H/a^3$ and $\dot H/a^3$, in order to assess $C^1$ (continuously differentiable) extendibility.
Strictly speaking, there are also subleading terms to Eq.~\eqref{eq:nablaRmunupp} of the form $\ddot H/a$ and $\dot H/a$
that could generally diverge as $a\rightarrow 0$. However, if $\ddot H/a^3$ and $\dot H/a^3$ are shown to be convergent,
then necessarily the quantities $\ddot H/a$ and $\dot H/a$ approach $0$ as $a\rightarrow 0$.

Equivalently, one may evaluate the covariant derivative of the Ricci tensor in the coordinate system defined in Sec.~\ref{sec:coordinate},
where $\mathbf{d}\lambda\equiv a\,\mathbf{d}t$. That way, the Ricci tensor is given by Eq.~\eqref{Riccilambda}, and its covariant
derivative is found to be
\begin{align}
 \nabla_\sigma & R_{\mu\nu}\mathbf{d}x^\sigma\otimes\mathbf{d}x^\mu\otimes\mathbf{d}x^\nu
 =-\frac{2(\ddot H-2\dot HH)}{a^3}(\mathbf{d}\lambda)^{\otimes 3} \nonumber \\
 &+\frac{2\dot HH}{a}g_{\alpha\beta}\left(\mathbf{d}x^\alpha\otimes\mathbf{d}x^\beta\otimes\mathbf{d}\lambda
 +\mathbf{d}x^\alpha\otimes\mathbf{d}\lambda\otimes\mathbf{d}x^\beta\right) \nonumber \\
 &+\frac{\ddot H+6\dot HH}{a}g_{\alpha\beta}\mathbf{d}\lambda\otimes\mathbf{d}x^\alpha\otimes\mathbf{d}x^\beta\,.
\end{align}
It is straightforward to see that the coefficient of the $\mathrm{d}\lambda^3$ term follows from evaluating $\partial_\lambda F$,
where $F\equiv-2\dot H/a^2$ is the coefficient of the $\mathrm{d}\lambda^2$ term in Eq.~\eqref{Riccilambda}.
For the second derivative of the form $\nabla_\omega\nabla_\sigma R_{\mu\nu}$, the most divergent component is the coefficient of
the $\mathrm{d}\lambda^4$ term, and it is given by $\partial_\lambda^2F$. In general, for the $r$-th covariant derivative, it is $\partial_\lambda^rF$.

In sum, we arrive at the following statement:
\begin{thm}
 A flat, asymptotically de Sitter, FLRW spacetime with boundary $\mathscr{B}^-$ at $t\rightarrow -\infty$
 (equivalently $a\rightarrow 0$ or $\lambda\rightarrow 0$)
 is $C^r$ inextensible and $\mathscr{B}^-$ is a $C^r$ p.\,p.\ curvature singularity if
 \begin{equation}
  \bigg|\lim_{\lambda\rightarrow 0}\frac{\partial^r}{\partial\lambda^r}\bigg(\frac{\dot H}{a^2}\bigg)\bigg|=\infty\,.
 \end{equation}
 Alternatively, $\mathscr{B}^-$ is not a $C^r$ p.\,p.\ curvature singularity and the spacetime is $C^r$ extendible
 if the quantity $\partial_\lambda^r(\dot H/a^2)$ is finite as $\lambda\rightarrow 0$.
\end{thm}

For $r=0$, this is the statement given at the end of Sec.~\ref{sec:pptetrad}. For $r=1$, one needs to evaluate the limit of
\begin{equation}
 \partial_\lambda\bigg(\frac{\dot H}{a^2}\bigg)=\frac{1}{a}\partial_t\bigg(\frac{\dot H}{a^2}\bigg)=\frac{\ddot H-2\dot HH}{a^3}\,,
\end{equation}
and so on. Alternatively, one can check that
\begin{equation}
 \lim_{\lambda\rightarrow 0}\partial_\lambda^r\bigg(\frac{\dot H}{a^2}\bigg)
 =H_\Lambda^r\lim_{a\rightarrow 0}\partial_a^r\bigg(\frac{\dot H}{a^2}\bigg)\,,
\end{equation}
provided the spacetime has already been shown to be $C^{r-1}$ extendible.
Interestingly, this implies that $C^\infty$ (infinitely differentiable) extendibility
is possible if, as $a\rightarrow 0$, $\dot H\sim a^q$ with $q\in\mathbb{Z}_{\geq 2}$. Indeed, in that case,
$\partial_a^r(a^{-2}\dot H)\sim a^{q-2-r}\rightarrow 0$ for $r<q-2$; $\partial_a^r(a^{-2}\dot H)\sim\mathrm{constant}$
for $r=q-2$; and $\partial_a^r(a^{-2}\dot H)=0$ for $r>q-2$.
If $\dot H\sim a^q$ with $q$ not an integer but still $q>2$, then the spacetime is at most $C^{\lfloor q\rfloor-2}$ extendible.
As expected, exact de Sitter space with $\dot H\equiv 0$ is another example of $C^\infty$ extendibility.

\bibliography{PEIU_final}

\end{document}